\documentclass{jfm}
\usepackage{graphicx}
\usepackage{epstopdf, epsfig}

\shorttitle{Competition evolution of Rayleigh-Taylor bubbles}
\shortauthor{Y.-S. Zhang, Z.-W. He. L. Li and B.-L. Tian}

\title{Competition evolution of Rayleigh-Taylor bubbles}

\author{You-sheng Zhang\aff{1}
 \corresp{\email{zhang\_yousheng@iapcm.ac.cn}}
  , Zhi-wei He\aff{1},Li Li\aff{1}
 \and Bao-lin Tian\aff{1}
  \corresp{\email{tian\_baolin@iapcm.ac.cn}} }
\affiliation{\aff{1}Institute of applied physics and computational mathematics, Beijing 100094, China
}

\begin{document}

\maketitle

\begin{abstract}
Material mixing induced by a Rayleigh-Taylor instability occurs ubiquitously in either nature or engineering when a light fluid pushes against a heavy fluid, accompanying with the formation and evolution of chaotic bubbles. Its general evolution involves two mechanisms: bubble-merge and bubble-competition. The former obeys a universal evolution law and has been well-studied, while the latter depends on many factors and has not been well-recognized. In this paper, we establish a theory for the latter to clarify and quantify the longstanding open question: the dependence of bubbles evolution on the dominant factors of arbitrary density ratio, broadband initial perturbations and various material properties (e.g., viscosity, miscibility, surface tensor). Evolution of the most important characteristic quantities, i.e., the diameter of dominant bubble $D$ and the height of bubble zone $h$, is derived: (i) the $D$ expands self-similarly with steady aspect ratio  $\beta  \equiv D/h \thickapprox (1{\rm{ + }}A)/4$, depending only on dimensionless density ratio $A$, and (ii) the $h$ grows quadratically with constant growth coefficient $\alpha  \equiv h/(Ag{t^2}) \thickapprox [2\phi/{\ln}(2{\eta _{\rm{0}}})]^2$, depending on both dimensionless initial perturbation amplitude ${\eta _{\rm{0}}}$ and material-property-associated linear growth rate ratio $\phi\equiv\Gamma_{actual}/\Gamma_{ideal}\leqslant1$. The theory successfully explains the continued puzzle about the widely varying $\alpha\in (0.02,0.12)$ in experiments and simulations, conducted at all value of $A \in (0,1)$ and widely varying value of ${\eta _{\rm{0}}} \in [{10^{ - 7}},{10^{ - 2}}]$ with different materials. The good agreement between theory and experiments implies that majority of actual mixing depends on initial perturbations and material properties, to which more attention should be paid in either natural or engineering problems.
\end{abstract}


\section{Introduction}

When two fluids are separated by an irregular perturbed interface and are accelerated in a direction opposite to that of the density gradient, Rayleigh-Taylor (RT) instability occurs and develops rapidly into the turbulent regime \citep{cheng2000dynamical} consisting of a bubble mixing zone (formed when a light fluid penetrates a heavy fluid) and a spike mixing zone (formed when a heavy fluid penetrates a light fluid). The mixing occurs ubiquitously in systems extending from micro to astrophysical scales \citep{Livescu2013numerical}.  As the simplest and primary descriptor of mixing, quantitative knowledge of the evolution of the structure and height of the mixing zone plays a fundamental role \citep{cheng2000dynamical, Dimonte2004,Dimonte2004dependence,Zhang2016evolution} for understanding many natural phenomena (e.g., supernova explosions) and engineering applications (e.g., inertial confinement fusion).

Up to now, it is well-known that the height of the bubble mixing zone $h$  grows quadratically with constant quadratic growth coefficient $\alpha  \equiv h/(Ag{t^2})$ \citep{Read1984experimental,george2002a,kadau2004nanohydrodynamics,Lim2010nonideal,youngs2017rayleigh}, and the diameter of the dominant bubble $D$  expands self-similarly with quasi-steady aspect ratio $\beta  \equiv D/h$  \citep{Alon1995power,Dimonte2000density, Dimonte2004}, where $g$  is acceleration,   $A \equiv (R - 1)/(R{\rm{ + }}1) \in [0,1]$ is dimensionless Atwood number defined with density ratio   $R \equiv {\rho _{heavy}}/{\rho _{light}}$. Due to the nearly stationary center of mass of mixing zone, the evolution of the spike mixing zone can be determined by that of the bubble mixing zone  \citep{cheng1999boundary, cheng2000density,Zhang2016evolution}. Consequently, knowledge of the values of $\alpha$ and $\beta$ becomes extremely important, but is still an open question \citep{Dimonte2000density,Dimonte2004,Dimonte2004dependence}. This puzzle may be attributed to the continued lack of a unified theory to regularise the observed $\alpha$ and $\beta$ with the dominant factors affecting mixing evolution, including the density ratio, initial perturbation amplitude and material properties (e.g., viscosity, surface tensor, miscibility or diffusivity) \citep{Read1984experimental,Linden1991Molecular,Dalziel1999Self,Dimonte2004dependence,kadau2004nanohydrodynamics,Ramaprabhu2005numerical,Mueschke2009Measurements,Banerjee20093D,Lim2010nonideal}.

In earlier studies, possibly influenced by the facts that all measured $\alpha \in (0.05,0.07)$ in different apparatus are independent of $A$ \citep{Read1984experimental,Youngs1989modelling,Kucherenko1991Experimental,Dimonte2000density}, researchers tended to find a universal $\alpha$  \citep{Alon1994scale,Alon1995power,Oron2001dimensionality}. However, except for the comparable results predicted with Front-Tracking method \citep{george2002a}, majority of shortwave-perturbation simulations over the past several decades predicted a much smaller $\alpha \approx 0.025$ \citep{Dimonte2004,dimonte2005recent,Cabot2006Reynolds,Youngs2013density,youngs2017rayleigh}. Moreover, a recent shortwave-perturbation experiment \citep{Olson2009Experimental} with miscible fluids indirectly validated previous numerical simulations \citep{Dimonte2004} and excluded the possibility of universal $\alpha$. Now the observed $\alpha$  has changed widely from 0.02 to 0.12 \citep{Dimonte2004,dimonte2005recent,Youngs2013density,youngs2017rayleigh}. Although many factors would affect the value of $\alpha$  and $\beta$  \citep{Dimonte2004}, we argue that the the major factors can be classified as two categories: (i) the initial perturbations at the interface and (ii) the material properties (e.g., density, viscosity, diffusivity, thermal diffusivity, surface tensor). This classification can be easily understood from the viewpoint of direct numerical simulation. The former determines the initial condition, and the latter determined the dimensionless parameters of governing equation, i.e. the Atwood, Reynolds, Schmidt, Prandtl and Weber Number \citep{Andrew2001Transition,george2002a}. However, up to now, a quantitative dependence of either $\alpha$ or $\beta$ on these factors has not been established.

In the other hand, now it is clear that self-similar evolution of RT-mixing can be achieved through two limiting and distinct mechanisms: bubble-merger and bubble-competition \citep{Dimonte2004dependence,dimonte2005recent,Youngs2013density}. If the interface is perturbed entirely by random combined waves with individual wavelengths $\lambda$ much shorter than the system width $L$ , bubbles will expand self-similarly via merging with their smaller neighbours \citep{Alon1994scale,Alon1995power}, leading to a universal lower bound $\alpha  \approx 0.025$  \citep{Dimonte2004,Youngs2013density,dimonte2005recent}. If perturbation involves some longer wavelengths $\lambda$ comparable to $L$ , the mixing at a later time evolves dominantly via the competition between the individual growth of the long waves seeded initially, leading to a larger $\alpha$ \citep{Dimonte2004dependence,Ramaprabhu2005numerical,dimonte2005recent,Youngs2013density}. In the latter situation, since the growth of individual wave closely relates to initial perturbation amplitude $h_0$ and material-property-associated linear growth rate $\Gamma$, the corresponding $\alpha$ may thus depend on dimensionless $\eta _0\equiv h_0/\lambda$ and $\phi \equiv\Gamma_{actual}/\Gamma_{ideal}$. Because the latter situation dominates in actual flow scenarios \citep{Haan1989onset,Dimonte2004dependence,Ramaprabhu2005numerical,Youngs2013density}, formulating  $\alpha (A,{\eta _0}, \phi)$ and $\beta (A,{\eta _0}, \phi)$  thus becomes extremely important but no self-consistent or satisfactory \citep{Zmitrenko1997evolution,Ramaprabhu2005numerical} theory has yet been established.

In this paper, a theory is established yielding analytic relations of  $\beta \approx (1{\rm{ + }}A)/4$ and  $\alpha\approx [2\phi/{\ln}(2{\eta _{\rm{0}}})]^2$, which successfully reproduce the observed results \citep{Read1984experimental,Youngs1989modelling,Kucherenko1991Experimental,Dimonte2000density, Ramaprabhu2005numerical,Youngs2013density} and formulate the disordered data \citep{Dimonte2004,dimonte2005recent}.

\section{Theory}

In this section, we present current theory by progressively clarifying the problems evolving from single-wave, wavepacket, and broadband-wave perturbations as follows. For the sake of conciseness, all mathematical derivations are given in Appendix.

\subsection{Evolution from single-wave perturbation}\label{problem1}
Up to now, until time $t_{Re}$ , corresponding to the possible appearance of a reacceleration stage, the development of an instability starting from a linear stage with exponentially growing $h(t)$  and transitioning into a quasi-steady stage with linearly growing $h(t)$  has been widely recognised and well formulated \citep{Zhang1998analytical,Mikaelian1998analytic,Sohn2003simple,Mikaelian2003explicit,Abarzhi2003rayleigh,Goncharov2002analytical,Ramaprabhu2005single,Zhang2016universality}. In the two-dimensional (2D) problem, \citet{Zhang2016universality} obtained a universal analytical expression  $h(t,A,{h_0},{\dot h_0})$ for an arbitrary $A$  and initial perturbation until $t_{Re}$ , where the dot and subscript $0$ denote, respectively, the derivative with respect to time and the value at the initial time. In the three-dimensional (3D) problem, one can obtain a similar expression by following Zhang's procedure \citep{Zhang2016universality} and by referring to previous analytical solutions \citep{Sohn2003simple,Mikaelian2003explicit} (see appendix \ref{appA}). The 3D solution also works until $t_{Re}$ , but its form is slightly complex. To simplify the solution, corresponding to the famous concept of boundary layer thickness ${\delta _{99}}$  introduced originally to divide the spatial-dependent velocity profile \citep{tani1977history} , we define a time boundary layer thickness $t_{99}$  to divide the time-dependent velocity evolution  $\dot h(t)$ as follows: at time $t_{99}$ , the instantaneous velocity asymptotically approaches the time-independent terminal velocity ${\dot h_\infty }$  of the quasi-steady stage, i.e., ${\dot h_{99}}{\rm{ = }}0.99{\dot h_\infty } \to {\dot h_\infty }$ . Consequently, for  $t \in [{t_{99}},{t_{{\mathop{\rm Re}\nolimits} }}]$, a linearly growing $h(t)$  is obtained (see appendix \ref{appB})
\begin{eqnarray}
h(t) =  - \lambda /\chi (A,{\eta _0}) + {\dot h_\infty }(A,g,\lambda )t,\;t \in [{t_{99}},{t_{{\mathop{\rm Re}\nolimits} }}],
\label{eq1}
\end{eqnarray}
where the dimensionless initial perturbation amplitude ${\eta _0}$  is very small in general problem \citep{Dimonte2004dependence,Ramaprabhu2005numerical,Youngs2013density}. Moreover, the introduction of $t_{99}$  directly leads to the following important findings (see appendix \ref{appB}): (a) the instability enters the quasi-steady stage when $h(t)$  grows to  ${h_{99}}({t_{99}}) \approx {C_2}\Theta \lambda$, (b) $t_{99}$  is universal scaled as  ${t_{99}} \propto \sqrt {\lambda \Theta /(Ag)}$, and (c) independent of  $A,\lambda ,g,{\eta _0},{\dot \eta _0}$ and $\Theta$,  ${\dot h_\infty }$ is always proportional to the average velocity  $\overline {\dot h_0^{99}}  \equiv ({h_{99}} - {h_0})/{t_{99}}$ as  ${\dot h_\infty }/\overline {\dot h_0^{99}}  = const = C$, where $\Theta(A)$  arises from potential theory and differs among different theories \citep{Goncharov2002analytical,Sohn2003simple} (see appendix \ref{appA}). These findings work for an arbitrary density ratio and initial perturbations and will be used below.

\subsection{Evolution from wavepacket perturbations}\label{problem2}
For an interface perturbed by a narrow wavepacket gathering around the dominant $\lambda_d$ , $h(t)$  evolves differently than problem \ref{problem1}. However, previous studies show that the root-mean-square amplitude ${h_{rms}} \equiv {{{[{{(2\pi )}^{ - 1}}{L^2}\int {h_k^{2}{k^{'}}dk} ]}^{1/2}}}$  grows similarly to that of $h(t)$ in problem \ref{problem1} until some autocorrelation time after $t_{99}$ \citep{Haan1989onset,Dimonte2004dependence}, where $k\equiv2\pi/\lambda$. Therefore, if we treat $h$  and $\lambda$  as $h_{rms}$  and $\lambda_d$ , respectively, equation (\ref{eq1}) still applies, and for conciseness hereafter this treatment is adopted. However, due to the introduction of a newly defined amplitude, the asymptotical velocity  $ {\dot h_\infty }$ in equation (\ref{eq1}) becomes an unknown and is determined, with the aid of the important findings emphasised in problem \ref{problem1}, as follows: (i) $ {\dot h_\infty }$  is determined by using finding (c), i.e., ${\dot h_\infty }{\rm{ = }}C\overline {\dot h_0^{99}}$, (ii) the constant  $C$ is viewed as a unique unknown parameter in the current theory and is determined with experimental data, (3)   $\overline {\dot h_0^{{\rm{99}}}}$ is determined by combining findings (a), (b) and the exponential growth of $h(t)$  in the linear stage, thus relating it to the initial perturbation $\eta_0$ and material-property-associated $\phi$ (see appendix \ref{appC}). After obtaining  $ {\dot h_\infty }$, the final solution derived from equation (\ref{eq1}) gives
\begin{eqnarray}
h(t) =  - \lambda /\chi (C,A,{\eta _0}) + Fr(C,A,{\eta _0},\phi)\sqrt {{{Ag\lambda }}/({{1 + A}})} t,\;t \in [{t_{99}},{t_{{\mathop{\rm Re}\nolimits} }}],
\label{eq2}
\end{eqnarray}
where $Fr$  (see appendix \ref{appC} for the expression) is called Froude number following the literature \citep{Dimonte2004dependence,Ramaprabhu2005numerical,dimonte2005recent}, and $\phi  \equiv \Gamma_{actual}/{\Gamma _{ideal}} \le 1$ is defined as the ratio of the actual linear growth rate $\Gamma_{actual} $  to the ideal linear growth rate  ${\Gamma _{ideal}} \equiv \sqrt {Akg}$, quantifying the relative decrement of the linear growth rate caused by various material properties \citep{Dimonte2004dependence}.

\subsection{Evolution from broadband wave perturbations}\label{problem3}
Considering an initial perturbation with arbitrary amplitude spectrum $h(\lambda)$ , the dimensionless $\eta_0(\lambda)$ , evaluated from $h_{rms}$  and $\lambda_d$  of the wavepacket, obviously depends on $\lambda$ . However, numerical studies \citep{Ramaprabhu2005numerical,Youngs2013density} show that the shape of the amplitude spectrum has little effect on either $\alpha$  or $\beta$ . Moreover, in most actual problems \citep{Haan1989onset,Dimonte2004dependence,Youngs2013density},  $h(\lambda ) \propto {\lambda ^2}/L$, for which spectrum the corresponding $\eta_0$  is independent of $\lambda$  \citep{Dimonte2004dependence}. Therefore, hereafter we only describe a theory for $\lambda$-independent $\eta_0$ .As for problems evolving from an interface perturbed by a broadband wave, the bubble-competition mechanism means that the bubble mixing zone grows as the competition of a series of amplitudes of individual wavepackets. Following Birkhoff and \citet{Dimonte2004dependence}, the self-similar evolution of the dominant bubble can be obtained by seeking the dominant wavelength $\lambda_d$  that maximises equation (\ref{eq2}). This is accomplished by requiring  $\partial h/\partial \lambda  = 0$ in equation (\ref{eq2}) to give  ${\lambda _d} = {\chi ^2}F{r^2}Ag{t^2}/[4(1 + A)]$, which when substituting into equation (\ref{eq2}) yields  $h = {\lambda _d}/\chi$. Naturally,  ${\alpha} \equiv h/(Ag{t^2}) = \chi F{r^2}/[4(1 + A)]$ , and  $\beta  \equiv D/h{\rm{ = }}\chi $ if we use the approximation relation ${\lambda _d} \approx D$  \citep{Goncharov2002analytical,Dimonte2004dependence,dimonte2005recent,Ramaprabhu2005single}. The unique unknown constant  $C \approx 8.843$ in $Fr(C)$  is determined with the observed \citep{Dimonte2000density}  $\beta [C,\Theta (A)] \approx 0.5$ at $A=1$, the only value of $A$  for which there is no disagreement \citep{dimonte2005recent} regarding the value of  $\Theta (A{\rm{ = }}1){\rm{ = }}1$ (see appendix \ref{appA} and \ref{appD}). As for  $\Theta(A)$, the analytical results $\Theta (A){\rm{ = }}2/(1 + A)$  \citep{Goncharov2002analytical} agrees very well with numerical simulations \citep{Dimonte2004dependence,dimonte2005recent,Ramaprabhu2005single} that are thus used in current theory. Based on the above results, we finally obtain (see appendix \ref{appD})
	 \begin{eqnarray}
\beta  \approx (1 + A)/4{\rm{,}}\quad {\alpha} \approx {[2\phi /\ln (2{\eta _0})]^2}.
\label{eq3}
\end{eqnarray}

\section{Discussions and Validations}

\begin{figure}
  \includegraphics[width=6.75cm]{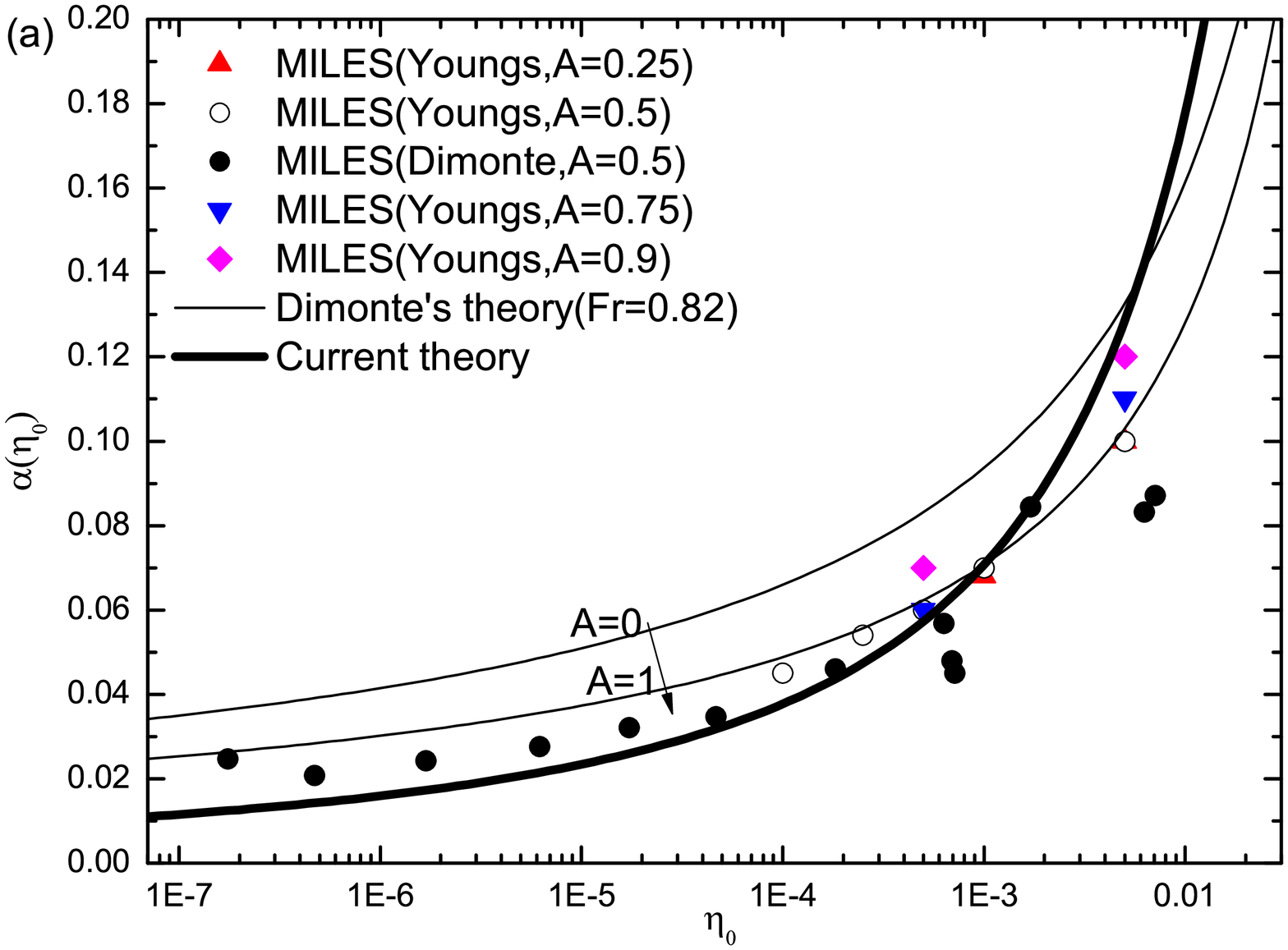}
  \includegraphics[width=6.5cm]{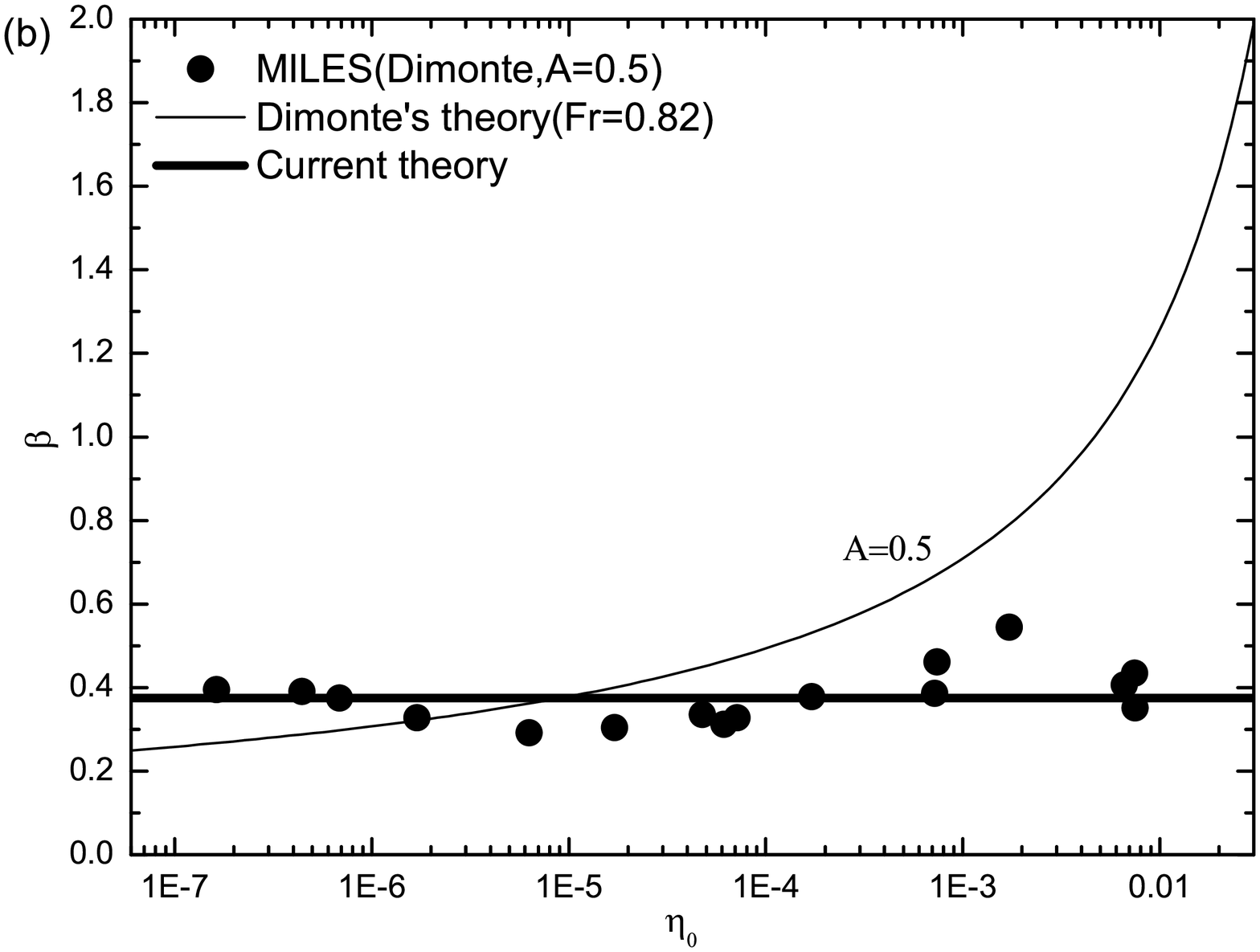}
  \caption{Comparison between theories and numerical simulations \citep{Ramaprabhu2005numerical,Youngs2013density}. According to the definitions, ${\eta _0} \approx \varepsilon$  and   ${\eta _0} = k{h_{rms}}/(2\pi)$, where  $\varepsilon$ and  $ k{h_{rms}}$ are the original dimensionless initial perturbation amplitudes given by \citet{Youngs2013density} and \citet{Ramaprabhu2005numerical}, respectively. }
\label{fig1}
\end{figure}

Equation (\ref{eq3}) predicts that (1) $\beta$ linearly depends on $A$, and only on $A$, (2) $\alpha$  depends on  both $\eta_0$ and $\phi$, while not on $A$. These predictions are distinct from Dimonte's prediction that either $\beta$ or $\alpha$  logarithmically depends on both $A$  and $\eta_0$ \citep{Dimonte2004dependence}, and is independent of material properties (i.e.,viscosity, surface tensor, miscibility). Compared with the familiar $\eta_0$ and $A$, the newly introduced $\phi$ is worth of some discussions. According to the definition of $\phi$, the ${\Gamma _{ideal}}$ denotes the linear growth rate of ideal fluids (i.e., without considering viscosity, surface tensor, miscibility or diffusivity). For the immiscible experiments \citep{Read1984experimental,Youngs1989modelling,Kucherenko1991Experimental,Dimonte2000density} conducted with $g\gg g_0$ (gravitation acceleration), the contribution of viscosity and surface tensor to the linear growth rate is neglectable \citep{Read1984experimental}, and thus we have ${\Gamma _{actual}}\rightarrow{\Gamma _{ideal}}$ and $\phi\thickapprox1$.  However, for immiscible experiment conducted with $g\sim g_0$ \citep{Linden1991Molecular}, the contribution of miscibility to the linear growth rate is considerable, and $\phi$ is smaller than 1. This may explain why the observed $\alpha$ in miscible experiment are smaller than that of immiscible experiments \citep{Youngs1989modelling,Kucherenko1991Experimental,Linden1991Molecular,Dimonte2000density}, provided that all the experiments involves comparable initial perturbation. Following this logic, we may explain why the $\alpha$ predicted by the immiscible Front-Tracking code is larger than that of predicted with miscible code \citep{george2002a}. Similarly, in simulations without involving extreme acceleration, the contribution of viscosity to linear growth rate is also considerable, and one must set $\phi<1$. Here we give an example to determine the $\phi$ in large eddy simulations \citep{Ramaprabhu2005numerical,Youngs2013density}, where either numerical or modeled viscosity is considerable. Following \citet{Dimonte2004dependence} viewpoint, the value of $\phi$ are dominantly determined by the monotonously increased $\phi({\lambda})$ from $\phi({\lambda _u})\thickapprox0.65$ \citep{Ramaprabhu2005numerical} to $\phi({\lambda _{\max }})=1$ \citep{Dimonte2004dependence}, where $\lambda_u$  and $\lambda_{max} $  denote, respectively, the most unstable and longest wavelength. Therefore, we set $\phi {\rm{ = (}}1 + 0.65{\rm{)/}}2{\rm{ = }}0.825$ by considering the simplest linear average.

Taking account of the discussions above, our current theory was validated systematically by reproducing all the available experiments and simulations, but only the series of ¡°Rocket Rig (RR)¡± \citep{Read1984experimental,Youngs1989modelling} and ¡°Linear Electronic Motor (LEM)¡± \citep{Dimonte2000density} experiments and the series of simulations using moderate \citep{Ramaprabhu2005numerical} and fine \citep{Youngs2013density} grids are presented here. In figures \ref{fig1} and \ref{fig2}, the numerical simulations and experiments are compared with current theory and Dimonte's theory. The Dimonte's predictions are plotted with $Fr=0.82$ to best fit the observed experimental data \citep{Dimonte2004dependence}. According to the above discussions, our predictions are plotted with $\phi=0.825$ and $\phi=1$ for simulations and experiments, respectively.

In figure \ref{fig1}, we validate current theory with a series of simulations. Dimonte predicted that $\alpha(A,\eta_0)$  depends on both $A$  and $\eta_0$ , so in the figure (1.a) we plotted the possible zone bounded by the two limiting curves with $A=0$ and $A=1$. However, as shown in this figure, scant data was located in this zone, and Youngs' simulation also do not support Dimonte's prediction that $\alpha$ decreases with increasing $A$ . In figure (1.b), a significant deviation of Dimonte's predictions from the simulation is observed. In fact, the simulations implied that $\alpha$  and $\beta$  depend, respectively, only on $\eta_0$  and $A$, which is consistent with our predictions.

In figure \ref{fig2}, we validate current theory with a series of experiments. For the LEM experiment, the current theory was exactly validated in the sense that equation (\ref{eq3}) predicted the observed $\alpha _{average}=0.05$ with the measured \citep{Dimonte2004dependence} $\eta _0^{average} = 4/(2\pi)  \times {10^{ - 4}}$ , giving  $\alpha _{scl}^{theory} = 1$. For RR experiments, due to the lack of $\eta_0$ , an averaged ${\eta _0} = 1.73 \times {10^{ - 4}}$ was estimated using the current theory to produce the measured ${\alpha _{average}} = 0.063$ . We also used this $\eta_0$  to check Dimonte's theory. As shown in the figure (2.a), the deviation of Dimonte's prediction from experiments increases with increasing $\eta_0$  and $A$, while our predictions agree very well. In figure (2.b), the solid and dashed thick lines denote, respectively, our predictions using the approximate relation $\lambda_d=D$  and exact relation $\lambda_d=1.07D$ \citep{Goncharov2002analytical,Dimonte2004dependence} from Goncharov's solution. As shown in this figure, the prediction with the exact relation agrees better.

\begin{figure}
\centering
\includegraphics[width=6.5cm]{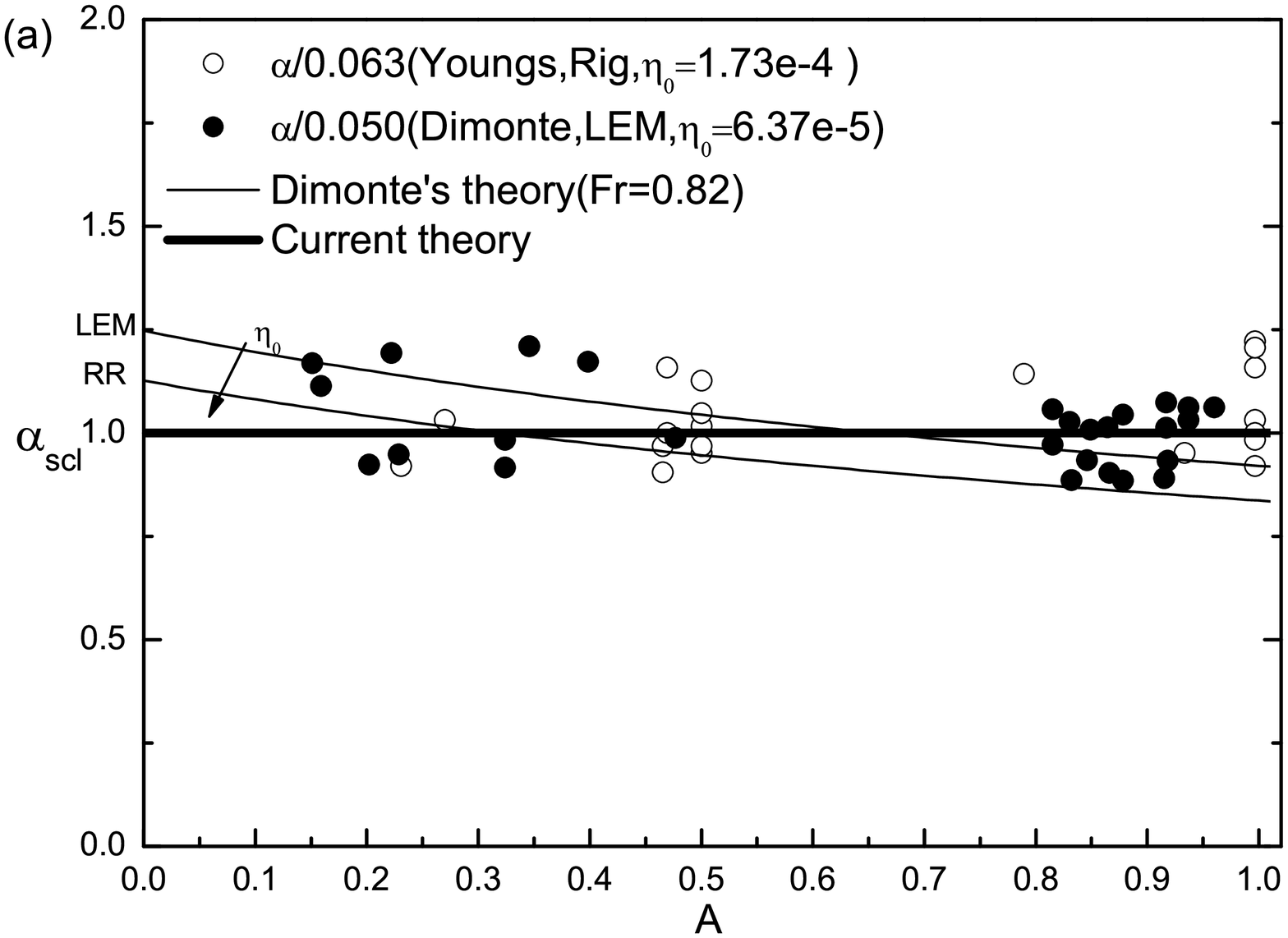}
\includegraphics[width=6.35cm]{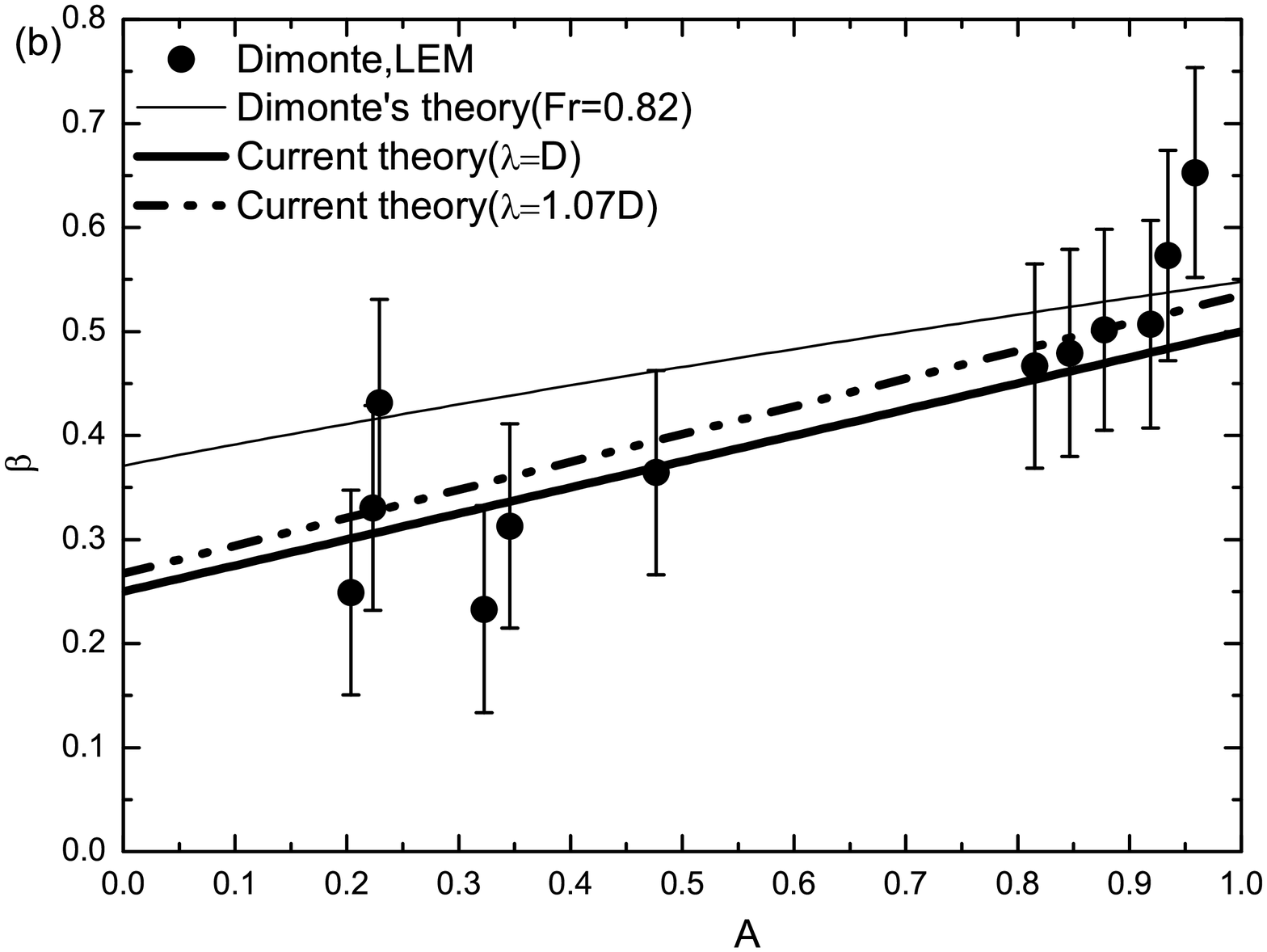}
\caption{ \label{fig2} Comparison between theories and experiments \citep{Read1984experimental,Youngs1989modelling,Dimonte2000density} for (a) scaled ${\alpha _{scl}} \equiv \alpha /{\alpha _{average}}$ and (b) $\beta$. In figure (a), regardless of the experiment or theory, $\alpha _{average}$  was set to 0.05 and 0.063 for LEM and RR data, respectively \citep{dimonte2005recent}.}
\label{fig2}
\end{figure}

From these comparisons, we can conclude that our predictions agree very well with both experiments and numerical simulations conducted at all density ratios $A \in (0,1)$  and widely varying ${\eta _0} \in ({10^{ - 7}},{10^{ - 2}})$ with different materials. Noting that we do not need to adjust parameters to fit experiments or simulations or to fit $\alpha$ or $\beta$ , so our reproduction is self-consistent.

\section{Conclusions}
Now it is clear that general evolution of turbulent RT mixing via two mechanisms: merging adjacent waves \citep{Alon1994scale,Alon1995power} and amplifying the individual waves presented in initial perturbations \citep{Dimonte2004dependence,Ramaprabhu2005numerical}.  The former is well studied previously and obeys a universal law \citep{Alon1994scale,Alon1995power}, and the latter is clarified in this paper and its dependence on density ratio, initial perturbation and material properties are formulated,too. The theoretical results are verified for all published results \citep{Read1984experimental,Youngs1989modelling,Dimonte2000density,Ramaprabhu2005numerical}. Our theory implies that most actual mixing depends on initial perturbations and evolves dominantly via the competition of individual wavepacket amplitudes, comprised of a narrow single-wave band. In addition, we also point out that in actual problems bubble-merge and bubble-competition exist simultaneously, so the two mechanisms should be considered simultaneously. However, this is beyond the scope of this paper, and will be addressed in another paper aiming to explain the observed transition \citep{Ramaprabhu2005numerical} at  ${\eta _0} \approx 5 \times {10^{{\rm{ - }}5}}$. We wish that current theory could promote the understanding of associated astronomical phenomena and the development of controllable thermonuclear fusion.

\section{Acknowledgements}
This work was supported from the Chinese Academy of Engineering Physics under Grant Number YZ2015015, and from National Nature Science Foundation of China under Grant Numbers 11502029, U1630138, 11572052, 11602028 and 11472059.

\appendix
\section{Analytical solution for $t \in [{t_0},{t_{{\mathop{\rm Re}\nolimits} }}]$}\label{appA}
Using Goncharov's \citep{Goncharov2002analytical} solution for the 3D problem, \citet{Mikaelian1998analytic,Mikaelian2003explicit} obtained an explicit $h(t)$  for a special initial condition until $t_{Re}$ . We noticed that Mikaelian's 3D solution has the same form as Zhang's 2D universal solution \citep{Zhang2016universality}. However, Zhang obtained a universal solution without requiring a special initial condition. Instead he used a reasonable \citep{Zhang1998analytical,Zhang2016universality} assumption that the curvature of the bubble tip is steady. Therefore, Mikaelian's special solution \citep{Mikaelian1998analytic,Mikaelian2003explicit} can be viewed as a natural consequence of the universal solution, as noted by \citet{Zhang1998analytical} for $A=1$. Based on this logic and the previous 3D analytical solution \citep{Goncharov2002analytical,Abarzhi2003rayleigh,Sohn2003simple}, we can write the 3D universal solution with the same form as the 2D universal solution as follows:
\begin{eqnarray}
\left\{ \begin{array}{l}
\ddot h = b(\dot h_\infty ^2 - {{\dot h}^2})/{{\dot h}_\infty }\\
\dot h = {{\dot h}_\infty }f(bt,\varepsilon )\quad \quad \quad \quad \quad \quad \quad \quad  \quad \quad  \quad \quad , t \in [{t_0},{t_{Re}}]\\
h = {h_0} + {b^{-1}}{{{{\dot h}_\infty }}}\ln [\cosh (bt) + \varepsilon \sinh (bt)]
\end{array} \right.,
\label{eq4}
\end{eqnarray}
where $b \equiv \sqrt {2{\beta _1}gA/(\lambda \Theta )}$ , ${\dot h_\infty } \equiv \sqrt {\Theta Ag\lambda /(2{\beta _1})}$, $f(bt,\varepsilon ) \equiv [\sinh (bt) + \varepsilon \cosh (bt)]/[cosh(bt) + \varepsilon \sinh (bt)] \ge f(bt,0) = \tanh (bt)$, and  $\varepsilon  \equiv {\dot h_0}/{\dot h_\infty }$ are very small in the general problem, ${\beta _1} \approx 3.832$  is the first zero of the Bessel function  $J_1(x)$, $\Theta(A)$  can be $2/(1+A)$ , 1 and so on by following Goncharov's \citep{Goncharov2002analytical}, Sohn's \citep{Sohn2003simple} and others¡¯ \citep{Abarzhi2003rayleigh} theories.

\section{Simplified linear solution for   $t \in [{t_{99}},{t_{{\mathop{\rm Re}\nolimits} }}]$}\label{appB}
Considering a special time $t = {t_{99}} = {C_1}/b$  with   ${C_1} = {\tanh ^{ - 1}}(0.99) = \ln \sqrt {199}$, we obtained   ${\dot h_{99}}({t_{99}}) \ge 0.99{\dot h_\infty }$ and   ${h_{99}}({t_{99}}) = ({\eta _0} + {C_2}\Theta )\lambda$ by substituting $t_{99}$  into equation (\ref{eq4}) and neglecting $\varepsilon$ , with   ${C_2} \approx \ln (\sqrt {199} /2)/(2{\beta _1})$. If we further ignore $\eta_0$ , we obtain ${h_{99}} \approx {C_2}\Theta \lambda$---finding (a). Furthermore, the above definition gives ${t_{99}} \propto \sqrt {\lambda \Theta /(gA)}$---finding (b). According to the definition, $\overline {\dot h_0^{99}}  \equiv ({h_{99}} - {h_0})/{t_{99}} = {C_2}\Theta \lambda /{t_{99}} = {\dot h_\infty }/C$  with $C \equiv {C_1}/(2{\beta _1}{C_2})$---finding (c). For $t>t_{99}$,$\dot h(t) \ge 0.99{\dot h_\infty }$  and tends to steady ${\dot h_\infty }$ , thus the evolution of $h(t)$  can be approximated (error   $\le 1\%$) as
\begin{eqnarray}
h = [{\eta _0} + \Theta {C_2}]\lambda  + {\dot h_\infty }(t - {t_{99}}) =- \lambda /\chi  + {\dot h_\infty }t, t \in [{t_{99}},{t_{{\mathop{\rm Re}\nolimits} }}],
\label{eq5}
\end{eqnarray}
where  $\chi  \equiv  - {[{\eta _0} + \Theta {C_2}(1 - C)]^{ - 1}}$ and the relation ${\dot h_\infty }{\rm{ = }}C\overline {\dot h_0^{99}} {\rm{ = }}C{C_2}\Theta \lambda /{t_{99}}$  is used.

\section{Determination of  $\overline {\dot h_0^{99}}$}\label{appC}
According to  $\overline {\dot h_0^{99}}  \equiv {C_2}\Theta \lambda /{t_{99}} $ derived in the appendix \ref{appB}, the key is to determine $t_{99}$ . Because $\Theta(A)$  differs in different theories except for $\Theta(A=1)=1$  \citep{dimonte2005recent} (see appendix \ref{appA}), we first determine $t_{99}$  for $A=1$ . For $A=1$ , ${h_{99}} \approx {C_2}\lambda  \approx \lambda /4$  (see the first equality of equation (\ref{eq5})), which is near the end of the linear stage. In the linear stage, $h(t) = {h_0}\cosh (\Gamma_{actual} t)$  \citep{Dimonte2004dependence}, where $\Gamma_{actual}  = \phi {\Gamma _{ideal}}$  is the actual linear growth rate. Thus, $t_{99}$  can be solved from the requirement that ${C_2}\lambda {\rm{ = }}{h_0}\cosh (\Gamma_{actual} {t_{99}})$  yielding ${t_{99}}(A) = \ln F({\eta _0})/(\phi \sqrt {2\pi } )\sqrt {\lambda \Theta (A)/(Ag)}$ , where  $F({\eta _0}){\rm{ = }}{C_2}/{\eta _0} + \sqrt {{{({C_2}/{\eta _0})}^2} - 1}  \approx 2{C_2}/{\eta _0}$. Noting that for $A=1$ the determined   ${t_{99}}(A{\rm{ = }}1) \propto \sqrt {\lambda \Theta /(Ag)}$ is self-consistent with the important finding (b), we thus assume that the above $t_{99}(A)$  applies for all $A$ . Thus,
\begin{eqnarray}
{\dot h_\infty }{\rm{ = }}C\overline {\dot h_0^{99}} {\rm{ = }}C{C_2}\Theta \lambda /{t_{99}}{\rm{ = }}Fr\sqrt {Ag\lambda /(1 + A)},
\label{eq7}
\end{eqnarray}
where $Fr \equiv C{C_2}\phi \sqrt {2\pi \Theta (1{\rm{ + }}A)} /\ln F({\eta _0})$  is the Froude number.

\section{Final results.}\label{appD}
Substituting $Fr$ and $\chi$  to the solution $\beta=\chi$  and ${\alpha} = \chi F{r^2}/[4(1 + A)]$  gives
\begin{eqnarray}
\left\{ \begin{array}{l}
\beta {\rm{ = }}\frac{1}{{\Theta {C_2}(C - 1){\rm{ - }}{\eta _0}}} \approx \frac{1}{{{C_2}(C - 1)\Theta }}\\
{\alpha}{\rm{ = }}\frac{{\pi \Theta }}{{2[{C_2}(C - 1)\Theta {\rm{ - }}{\eta _0}]}}\frac{{{{(C{C_2}\phi )}^{2}}}}{{{{\ln }^{2}}F({\eta _0})}} \approx \frac{{\pi {C_2}{C^2}{\phi ^{2}}/(2C - 2)}}{{{{\ln }^{2}}[{\eta _0}/(2{C_2})]}}
\end{array} \right.,
\label{eq8}
\end{eqnarray}
where the last relation is obtained by neglecting $\eta_0$  when it is a small quantity. When $A=1$ , $\Theta(1)=1$ , and the constant $C\approx 8.843$  is determined to meet the observed \citep{Dimonte2000density} quantity $\beta(A=1)\approx0.5$ . Substituting $C\approx 8.843$  and $\Theta=2/(1+A)$  into equation (\ref{eq8}) immediately gives equation (\ref{eq3}).

\bibliographystyle{jfm}
\bibliography{jfm-instructions}

\end{document}